# A limit for the values of the *Dst* geomagnetic index

**F.J. Acero[1,2], J.M. Vaquero[2,3], M.C. Gallego[1,2], and J.A. García[1,2]**

1 Departamento de Física, Universidad de Extremadura, E-06071 Badajoz, Spain; fjacero@unex.es

2 Instituto Universitario de Investigación del Agua, Cambio Climático y Sostenibilidad (IACYS), Universidad de Extremadura, E-06006 Badajoz, Spain

3 Departamento de Física, Universidad de Extremadura, E-06800 Mérida, Spain

Corresponding author: F. J. Acero (fjacero@unex.es)

**Key points**: Extreme Value Theory, *Dst* geomagnetic index, geomagnetic storm

Abstract

The study of the extreme weather space events is important for a technological dependent society. Extreme Value Theory could be decisive to characterize those extreme events in order to have the knowledge to make decisions in technological, economic and social matters, in all fields with possible impacts. In this work, the hourly values of the *Dst* geomagnetic index has been studied for the period 1957-2014 using the peaks-over-threshold technique. The shape parameter obtained from the fit of the generalized Pareto distribution to the extreme values of the |*Dst*| index leads to a negative value implying an upper bound for this time series. This result is relevant, because the estimation of this limit for the extreme values lead to 850 nT as the highest expected value for this geomagnetic index. Thus, from the previous characterization of the Carrington geomagnetic storm and our results, it could be considered the worst case scenario.

## 1. Introduction

Although extreme space weather events are rare, their study has a huge interest in the possible substantial impact on our society that is technologically dependent (NRC, 2008). Although such events have only occasionally been observed, we have a basic picture of the factors necessary for them to be produced. However, some questions about them remain unanswered. In particular we know little about the limits and about the maximum size of such events (Riley et al. 2018).

In this contribution, we have made a statistical study of the *Dst* geomagnetic index, which has been widely used to describe the size of extreme geomagnetic storms. For this, we use a set of statistical techniques commonly called Extreme Value Theory (EVT). These techniques are beginning to be used to better characterize the extreme values of different indices related to space storms (Acero et al. 2017, 2018; Elvidge and Angling 2018).

2. Data and methodology

The *Dst* index represents the hourly average disturbance of the geomagnetic field in the Earth's low-latitude region and is a measure of the ring current intensity. For this study, hourly values of the *Dst* are used. The study period spans from 1957 to 2014 corresponding to the final database of the *Dst* index. This database is available from the World Data Center for Geomagnetism corresponding to the Kyoto University in Japan (http://wdc.kugi.kyoto-u.ac.jp/index.html). Data for 2015 and 2016 are also available but with a provisional character and were not considered in this study.

A branch of the statistics, widely used to assess the probability of rare and extreme events is the Extreme Value Theory (EVT). This branch deals with the stochastic behavior of the extreme values in a process and provides the solid fundamentals needed for the statistical modelling of such events and the computation of extreme risk measures. There are different approaches in the EVT. The Peaks-over-threshold (POT) is the one chosen to develop this study. In this technique, a high enough threshold is defined and the exceedances over this threshold are fitted to a Generalized Pareto Distribution (GPD). In the asymptotic limit for sufficiently large thresholds, with the observed geomagnetic index $Dst(t)$, the distribution of independent exceedances $X_u(t) = Dst(t) - u$ with $Dst(t) > u$ follows a GPD given by

$$P(X < x) = 1 - \left(1 + \frac{\xi \cdot x}{\sigma}\right)^{-1/\xi} \tag{1}$$

with $x > 0$ and $1+\xi\, x/\sigma > 0$, where $\sigma$ is the scale parameter, and $\xi$ is the shape parameter ($\xi \neq 0$). Negative shape parameter values indicate that the distribution has an upper bound, while values positive or zero values indicate that the distribution has no upper limit (Coles 2001). The scale parameter $\sigma$ gives information about the variability of the distribution.

To apply the POT approach it is necessary to define the threshold ($u$). There are different techniques to find the best threshold for a dataset. As the extreme values of *Dst* are negative, its absolute value will be considered. Figure 1 shows the results obtained through applying the methodologies for the threshold selection. Top panel shows the mean residual life plot and the bottom panel, the fit of the generalized Pareto distribution parameters at a range of threshold. Firstly, looking at the mean residual life plot, there is some evidence for linearity above u ~ 250 nT as marked by the beginning of the red line and, besides, the 95% confidence interval enlarges from that value. Secondly, the objective of using the fit of the GPD to a range of threshold is to look for the stability of the parameter estimates, being the reparametrized scale (σ*) and the shape parameter (ξ) near-constant above u ~ 250 nT as can be seen in the two bottom panels of Figure 1. Then, for the study period 1957-2014, the proposed threshold to study extreme storms using the EVT is |*Dst*|=*u*=250 nT. This value coincides with the used in Gonzalez et al. 2011 to define superintense geomagnetic storms (*Dst* ≤ -250 nT).

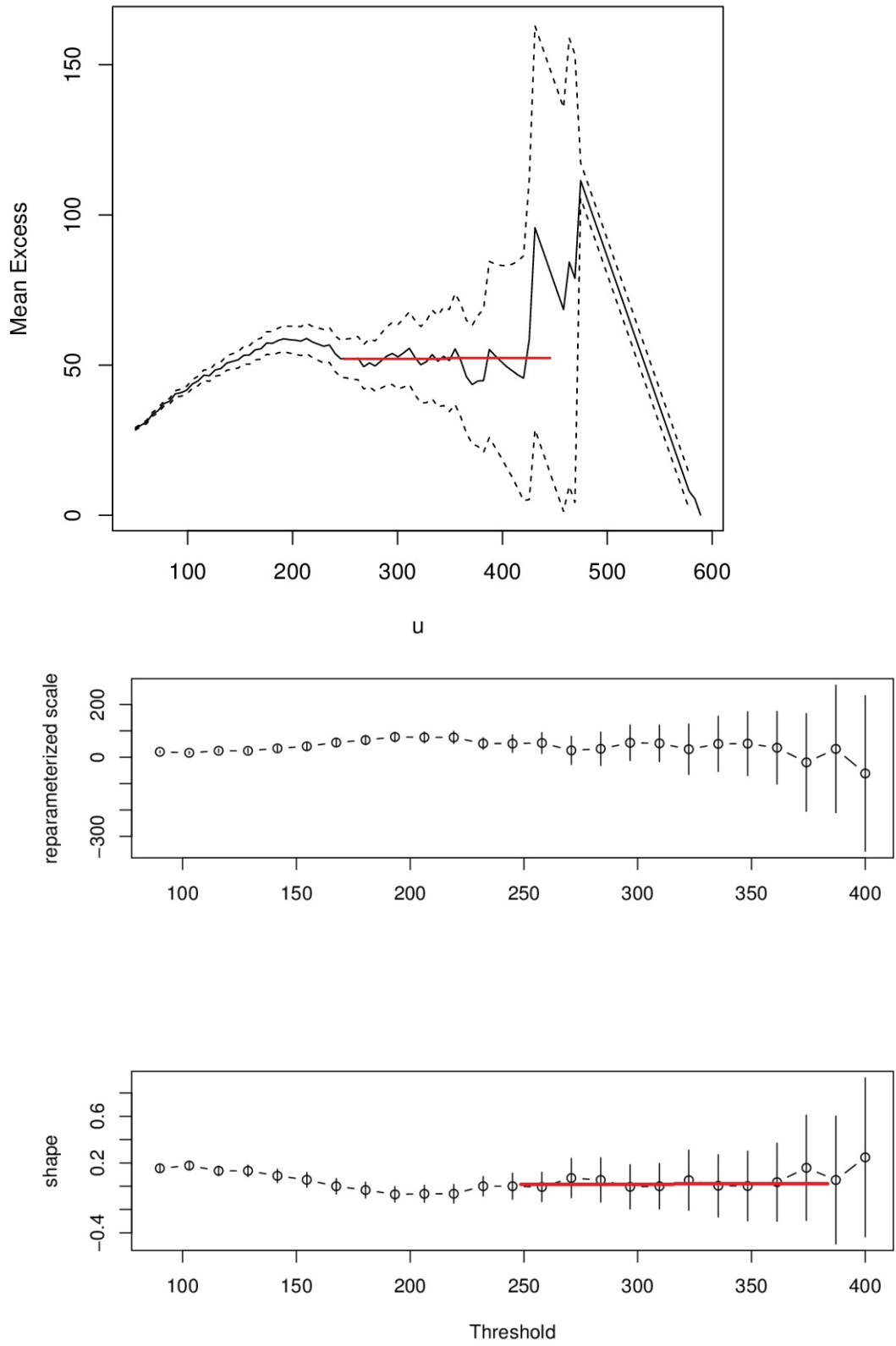

Figure 1. Mean residual life plot (top) and parameter estimates against threshold (two bottom panels) for *Dst* time series. Red line shows the linear stretch of the estimates.

Once the threshold was chosen, the EVT requires the observations fitted to the GPD analysis to be independent. As hourly data are considered, two or more consecutive hours could be under the chosen threshold being the exceedances grouped in clusters. Then, it is necessary a declustering procedure in order to assure the independence of the extreme observations. The procedure chosen is named 'run declustering' and consists in choosing a run length, *r*, and establishes that any extreme observations separated by fewer than r non-extreme observations belong to the same cluster. In the present study, *r*=48 hours was chosen in order to select the extreme observation for each cluster. The same value for the run length was used in previous studies about extreme values of the *Dst* (i.e., Tsobouchi and Omura 2007). In order to illustrate the behavior of a storm and to better understand the necessity of the declustering, Figure 2 shows the *Dst* values for the highest recorded storm between the 13th and 14th March 1989. In that event, from all the exceedances (*Dst* < -250 nT), only the highest one (*Dst* = -589 nT) was taken for the analysis to assure the independence of the extreme events. The same procedure is used for all the storms with *Dst* below the chosen threshold leading to 38 independent clusters that were fitted to a GPD in order to study the behavior of these extreme events.

The parameters of the GP distribution (shape and scale), were estimated by the maximum likelihood (ML) method using the in2extRemes statistical R software package for extreme values (Gilleland & Katz 2013). Once the parameters had been estimated, the confidence interval (CI) for each parameter was evaluated by a bootstrap procedure using 500 replicates (Gilleland & Katz 2013).

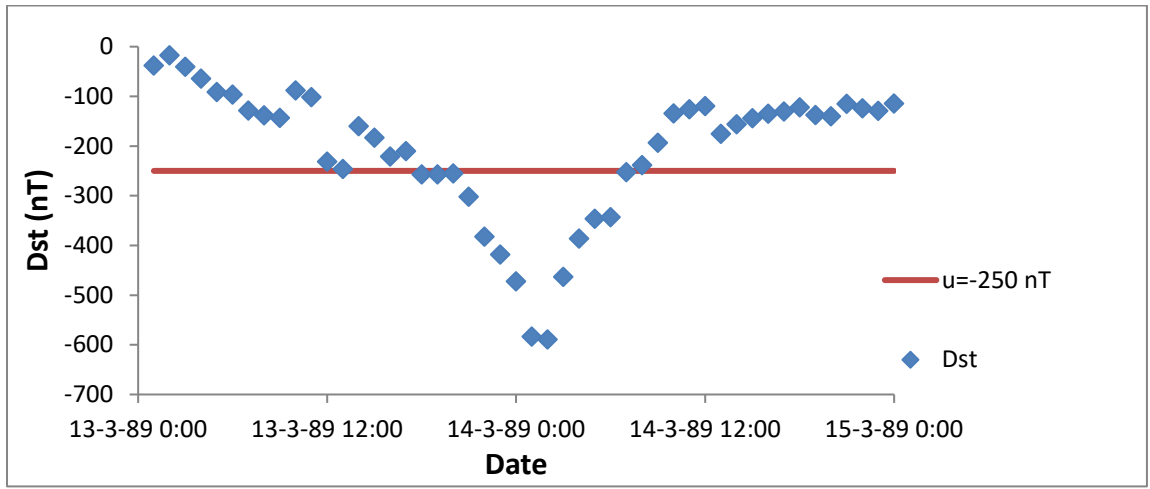

Figure 2. Hourly *Dst* values (points) between 13 and 15 March 1989, showing the event with the highest value of *Dst*. The red line shows the chosen threshold. The x-axis format is day-month-year hour:minute (UT time).

3. Results

Once the threshold was selected and the declustering the procedure applied, 38 independent superintense (as named by González et al. 2011) storms for the observed period were obtained. The distribution of these storms is irregular as can be seen in both, the temporal evolution and the histogram in Figure 3. There are 3 years with 4 events per year, another 3 years with 3 events per year, 4 years with 2 events per year, and 9 years showing 1 event. Then, only 19 of the 57 total years show storms and there are periods with several consecutive years showing none superintense storms, the largest covering from 1971 to 1980.

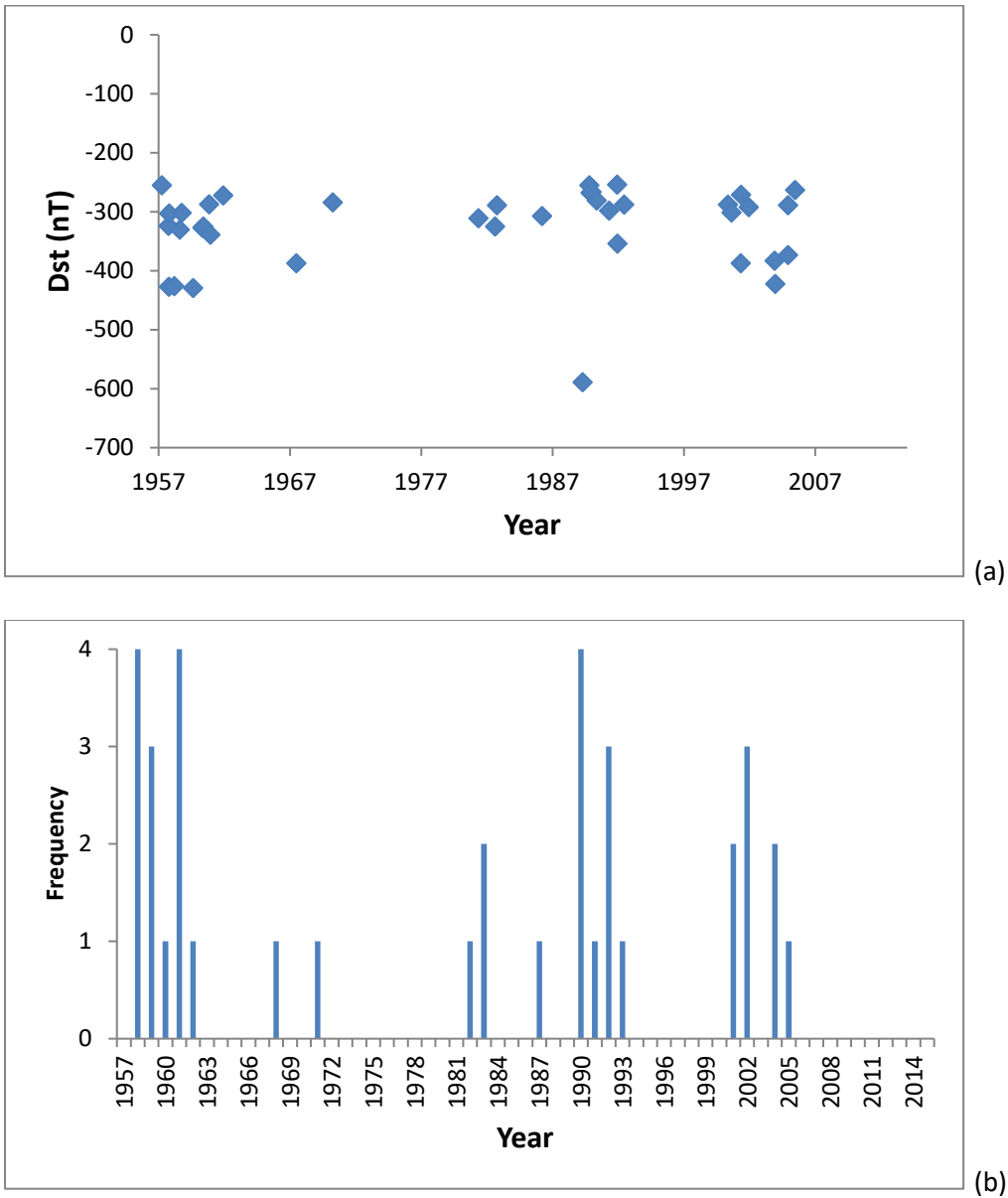

Figure 3. Temporal evolution (a) and histogram (b) of the chosen independent extreme events.

In order to assess the accuracy of the GPD model fitted to the independent extreme values of the *Dst*, different diagnostic plots were used and are shown in Figure 4. Both the probability plot and the quantile plot confirm the validity of the fitted model because both set of plotted points are near-linear. Only the highest value (|*Dst*|=589 nT) in the quantile plot seems an outlier, due to its prominent difference with the following four superintense storms with values in the interval |*Dst*|= (429 nT, 425 nT). Besides, as observed in the density plot in Figure 4, the density estimate seems consistent with the histogram of the data.

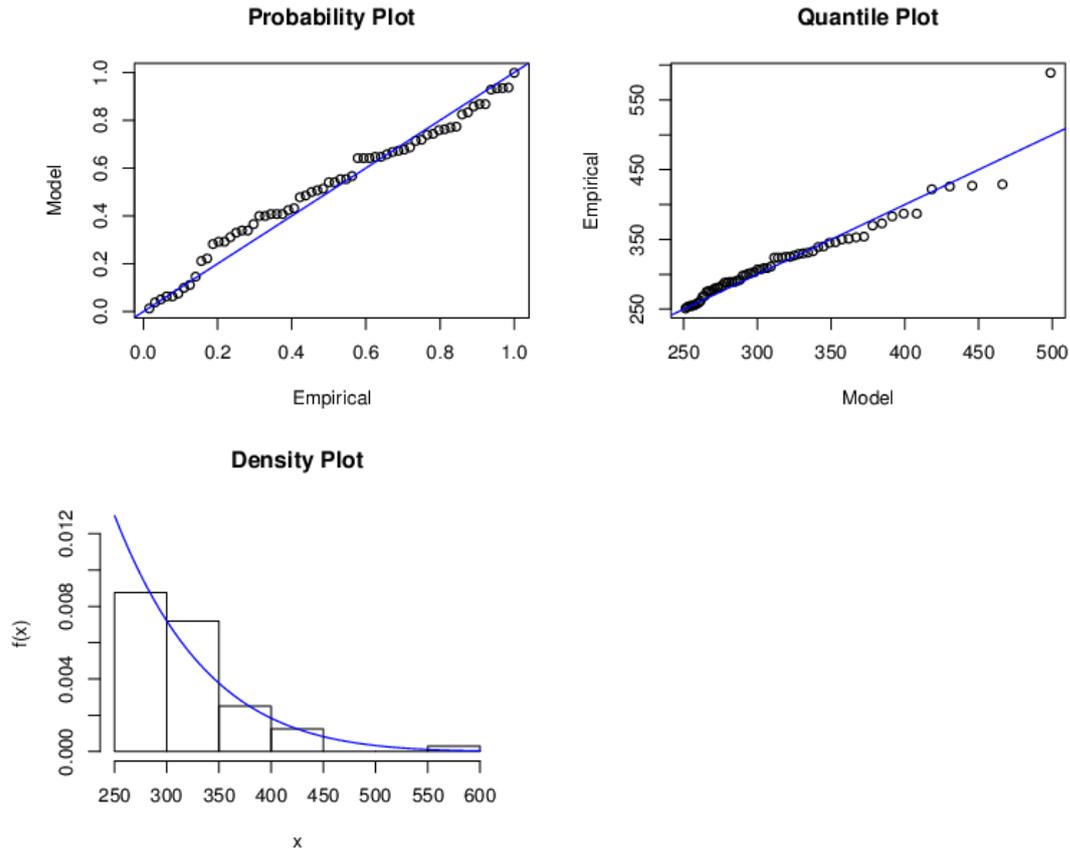

Figure 4. Diagnostic plots from fitting a GPD to the maximum *Dst*. Plots are a QQ-plot of empirical data quantiles against GP fit quantiles (top left panel), QQ-plot of randomly generated data from the fitted GP against the empirical data quantiles (top right panel), and (bottom panel) empirical density of the observed maxima *Dst* (histogram) with GP fit density (dark blue line).

Once the model has been tested and validated, table 1 lists the results for the estimated GPD parameters with the corresponding 95% CI obtained by a bootstrap procedure. The scale parameter ($\sigma$) gives information about the variability, and the shape parameter ($\xi$) about the shape of the extreme independent events distribution. The shape parameter and its CI is nearly always negative, meaning the distribution has an upper bound. This result is different than the shape parameter obtained by Tsobouchi et al. (2007) ($\xi$= 0.177) using as threshold $u$=|$Dst$|=280 nT and as study period 1957-2001. The main reason to obtain such different results, as well as the use of a shorter period, is that Tsobouchi et al. (2007) did not use a declustering procedure. Therefore, the assumption of the EVT to use independent observations was not verified.

Table 1. Estimates of the GPD parameters and their 95% confidence intervals obtained by the delta method.

| Scale parameter ($\sigma$)  [95% CI] | Shape parameter ($\xi$) [95% CI] |
|---|---|
| 76.96 [54.30, 99.61] | -0.13 [-0.29, 0.04] |

As established by the EVT, if $\xi<0$ the distribution of excesses has an upper bound u-σ*/ξ (Coles 2001). The estimation of this bound leads to a value of |*Dst*|=851.02 nT [183.13, 1518.91], being the limit of the exceedances. The confidence interval (CI) estimated with the delta method (Coles, 2001) is wide because the GPD parameters also show high CI. Despite of the selected threshold is appropriate in the sense of the EVT, with the purpose of confirming the estimated bound, two new values for *u* in the neighborhood of *u*=250 were checked. Thus, for *u*=240 and 260 nT, Table 2 shows the results for the shape parameter, that is also negative and, therefore, the bound for the |*Dst*| index.  All the estimates for the bound show values between 800 and 850 nT and considering the confidence intervals, none of the three values reach the estimated value of -1760 nT by Tsurutani et al. (2003).

Table 2. Estimates of the shape parameter, the |*Dst*| bound, and their 95% confidence intervals obtained using the delta method for different thresholds.

| Threshold (nT) | Shape parameter ($\xi$) [95% CI] | |*Dst*| Bound  [95% CI] |
|---|---|---|
| u=240 | -0.15 [-0.30, 0.01] | 802.16 [311.99, 1292.34] |
| u=250 | -0.13 [-0.31, 0.04 | 838.17 [192.64, 1483.70] |

The return level (RL) is a procedure usually used to interpret extreme events of any variable in order to estimate the amplitude of rare events expected to be exceeded on average once every N years. Three different values of N were selected, N= 50, 100, and 200, the first (second and third) corresponding to N shorter (longer) than the study period (58 years). Table 3 lists estimate of the RLs for the three values of N chosen. It is remarkable that for N= 100 years, longer than the observed period, the RL obtained is lower than the highest observed value of *Dst*=-589 nT. The RL plot is shown in Figure 5. Two facts are highlighted from the plot: firstly, the highest observed value is out of the 95% CI of the RL plot being a really special event in the context of the remaining superintense geomagnetic storms, and secondly, the RL plot decreases as the return period increase leading to an interesting result: the probability of reaching *Dst* values higher than the observed ones in the near future is scarce. Then, in line with the quantile plot previously showed (Figure 4) the highest observed value is even out of the 95% confidence interval of the return level plot. It seems one of the exceptional values ever registered for the geomagnetic activity.

Table 3. Different estimates of the return level and their 95% confidence intervals obtained by bootstrapping

| 50-year RL  [95% CI] | -558.37 [-648.92; -467.83] |
|---|---|
| 100-year RL  [95% CI] | -576.90 [-680.44; -473.37] |
| 200-year RL  [95% CI] | -600.18 [-722.47; -477.90] |

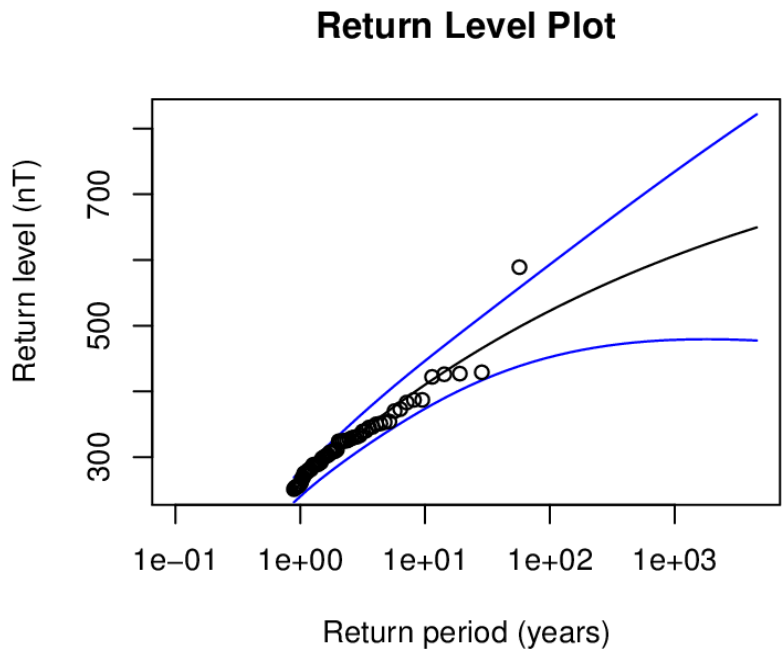

Figure 5. Return level plot (log scale) for maximum *Dst* with 95% normal approximation pointwise confidence intervals (blue line).

### 4. Conclusions

In this work, we have applied the POT technique, in the context of the Extreme Value Theory, to hourly average disturbance of the geomagnetic field using the *Dst* for the period 1957-2014. The shape and the scale parameter for the generalized Pareto distribution, which models the probability distribution of the exceedances, were estimated. A study for the threshold selection lead to |*Dst*|=250 nT as an appropriate

threshold in the sense of the EVT. The negative value obtained for the shape parameter reveals the existence of an upper limit of the distribution of excesses. This means a value that is not likely to be exceeded but with the limitation of the study period used in this work. Besides, to interpret the extreme values, the N-year return levels were estimated for periods shorter (N=50) and longer (N=100, 200) than the observed period (58 years). The 50- and 100-year (200-year) RLs show values under (over) the highest observed values. These results suggest that the highest extreme values of the time series for the geomagnetic activity has been reached in the past and are not expected to be exceeded in the nearest future. Moreover, the return levels are relaxing as the value of N increase, and it will be unlikely that the geomagnetic index *Dst* will attain values greater than the already observed ones in the future.

The existence of an upper limit for the geomagnetic index |*Dst*| with an approximate value of -851 nT conflicts with some previous estimations of the most intense geomagnetic storm, the Carrington's storms in 1859, with an estimated value of -1760 nT by Tsurutani et al. (2003). However, our estimation agrees with the *Dst* value associated with the Carrington's storm provided by Siscoe et al. (2006), also approximately -850 nT. Thus, the Carrington geomagnetic storm could be considered the worst case scenario according to our results. Despite the wide possible range of values of our estimation, this coincides with the latest *Dst* values that have been estimated for the Carrington storm around -900 nT with a range from -850 to -1050 given by the empirical determinations (Li et al. 2006; González et al. 2011; Cliver and Dietrich, 2013).

In any case, although the theoretical and empirical values of the extreme value of the *Dst* index coincide, we must bear in mind that solar and geomagnetic storms are very complex phenomena. Therefore, the characterization of these phenomena with a single index shows us the great phenomenological variety presented by extreme magnetospheric events (Riley et al. 2018).

Acknowledgements

This research was supported by the Economy and Infrastructure Counselling of the Junta of Extremadura through project IB16127 and grant GR15137 (co-financed by the